\documentclass[dvips]{article}

\usepackage{url}
\usepackage{icrc2011}

\title{Discovery of VHE gamma-ray emission from the SNR G\,15.4+0.1 with H.E.S.S.}

\newcommand{\etal}{\MakeLowercase{\textit{et al. }}} 

\newcommand{\g}{G\,15.4$+$0.1}
\newcommand{\hessj}{HESS\,J1818$-$154}

\shorttitle{P. Hofverberg \etal VHE $\gamma$-ray emission from G15.4+0.1}

\authors{P. Hofverberg$^{1\star}$, R.C.G. Chaves$^{1}$, J. M\'ehault$^{2}$ and M. de Naurois$^{3}$, for the H.E.S.S. Collaboration }
\afiliations{
$^1$Max-Planck-Institut f\"ur Kernphysik, P.O. Box 103980, D 69029 Heidelberg, Germany\\
$^2$Laboratoire de Physique Th\'eorique et Astroparticules, Universit\'e Montpellier 2, CNRS/IN2P3, CC 70, Place Eug\`ene Bataillon, F-34095 Montpellier Cedex 5, France\\
$^3$Laboratoire Leprince-Ringuet, Ecole Polytechnique, CNRS/IN2P3, F-91128 Palaiseau, France\\
}
\email{$^{\star}$petter.hofverberg@mpi-hd.mpg.de}

\abstract{Supernova remnants (SNRs) have emerged as one of the largest source classes in very-high-energy (VHE; E$>$0.1\,TeV) astronomy. Many of the now known VHE $\gamma$-ray emitting SNRs have been discovered by the H.E.S.S. imaging Cherenkov telescope array, thanks to its unique access to the inner galaxy. Statistically-significant emission of VHE $\gamma$-rays has now been detected from the direction of the supernova remnant \g. While the centroids of the H.E.S.S. source and the shell-type SNR are compatible, the VHE morphology suggests a center-dominated source at TeV energies, something which is at odds with the shell-like morphology observed at radio frequencies. This suggests that H.E.S.S. may be observing TeV emission from a previously unknown pulsar wind nebula (PWN) located within the boundaries of the radio shell. If this interpretation is correct, \g\ would in fact be a composite SNR, the first case in which an SNR is identified as a composite on the basis of VHE $\gamma$-ray observations. Archival data from MAGPIS gives exciting hints that there is radio emission from the central parts of the remnant, giving support to this hypothesis. Unfortunately, image artefacts from a nearby strong radio source produce considerable uncertainties in the radio analysis. Additional observations in both the radio and X-ray are needed to confirm the composite nature of \g\ suggested by H.E.S.S.}

\keywords{Observations: VHE $\gamma$-rays, ISM: Individual: \g}

\begin{document}
\maketitle

\section{Introduction}
Very-high-energy (VHE; E$>$0.1\,TeV) $\gamma$-ray emission has now been detected from more than 60 Galactic sources,
and the vast majority of these are in some way connected to supernova remnants (SNRs) \cite{Gast2011}.
TeV $\gamma$-rays have been observed from SNR shells themselves and also from the pulsar wind nebulae (PWNe) 
at the center of many SNRs. While the former gives rise to VHE $\gamma$-ray emission tracing the shock-front of the SNR,
the latter shows a centre-filled morphology, i.e. a central nebula, generally visible from low-energy radio
wavelengths up to $\gamma$-ray photon energies of tens of TeV in the most extreme cases. 
Young PWNe typically have a rather symmetric morphology and are centrally located inside the host SNR.
Such PWNe show a distinct spectral fingerprint with a roughly flat radio spectral index and a much steeper spectrum at
higher energies \cite{pwnevolution}.
The pulsar powering the nebula is usually found close to its birthplace at the centre of the SNR in such systems. 
However, only in a few cases has the pulsar actually been detected. 
One prominent example of a young PWN is found in the SNR G\,0.9$+$0.1 which was first classified as a composite SNR on
the basis of observations in radio \cite{helfland1987} then subsequently detected in TeV $\gamma$-rays \cite{hessg0.9},
before ultimately the energetic pulsar was found in a deep radio observation of the source \cite{radiog0.9}.


The SNR \g\ is a poorly studied object which was initially discovered in a 90-cm survey of the inner Galaxy
conducted by the Very Large Array (VLA) \cite{brogan2006}. The SNR was reported to have a shell-like morphology with a size about $14^{\prime} \times 15^{\prime}$ and an average spectral index $\alpha = -0.6 \pm 0.2$, indicating that
the radio emission is dominated by non-thermal synchrotron emission from the shell. For the reasons outlined above, 
this source is a prime candidate for H.E.S.S. observations. Furthermore, since \g\ is larger than the H.E.S.S. point-spread-function
(PSF; $\sim6^{\prime}$), morphological studies with H.E.S.S. have the capability to distinguish between a shell-type
SNR or PWN-like origin of the VHE $\gamma$-ray emission, thus shedding more light on this little known SNR.

\section{H.E.S.S. observations and results}
The region around \g\ was observed with H.E.S.S. between 2004 and 2010; 
the dataset consists primarily of observations from the H.E.S.S. Galactic Plane Survey \cite{Gast2011} and offset observations 
of nearby sources.
The dataset in this region has a live-time of $\sim$145\,h after standard H.E.S.S. quality selection \cite{hesscrab}, 
although the effective live-time at the target in question is significantly lower due to the large average offset of the
pointings from the target. 
The dataset was analyzed using the standard H.E.S.S. analysis \cite{hesscrab} which employs the Hillas second moment method
\cite{hillas1985} for discriminating between $\gamma$-ray- and hadron-induced extensive air showers (EASs).
\emph{Hard cuts}, where at least 200 photoelectrons are required in each recorded EAS image,
were used instead of \emph{standard cuts} (80 p.e.) in this analysis, in part because they give the improved angular resolution
which is crucial for this analysis.
To generate 2D images, the \emph{ring background method} \cite{ringbkg} was used with an inner radius of $0.7^{\circ}$.
All presented H.E.S.S. results have been cross-checked with an alternative analysis chain using an independent calibration and
$\gamma$-ray/hadron separation \cite{modelpp}.

\subsection{Detection and morphology}
\begin{figure}[!t]
  \vspace{5mm}
  \centering
  \includegraphics[width=3.25in]{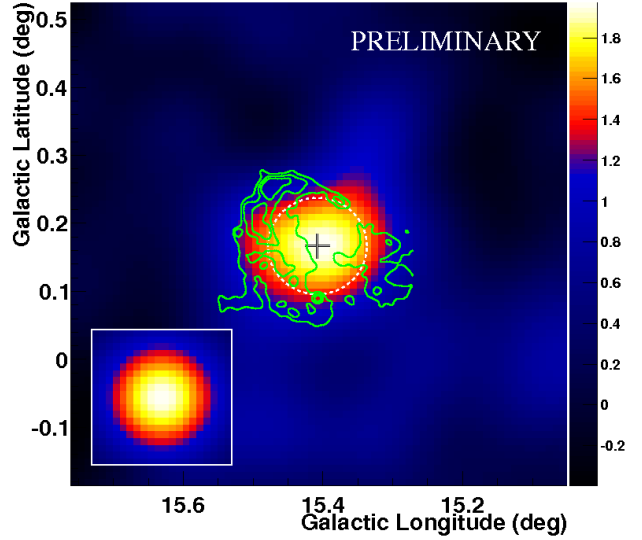}
  \caption{
The VHE $\gamma$-ray excess (in units of $\gamma$-rays arcmin$^{-2}$) around SNR \g. 
The image has been corrected for the varying exposure across the field-of-view and 
smoothed with a 2D Gaussian with a width of $0.06^{\circ}$.
The H.E.S.S. PSF for this analysis is shown in the bottom left inset.
The color scale is chosen such that the blue-red transition occurs at roughly 4-$\sigma$ significance.
The best-fit centroid of the $\gamma$-ray excess is indicated by a cross, the size of which
corresponds to the sum of both statistical and systematic uncertainties.
The best-fit extension of \hessj\ is represented as a circle.
The green contours show the intensity (at 0.0175, 0.035 and 0.0525\,Jy\,beam$^{-1}$) of the radio emission from the SNR shell,
based on 90-cm VLA observations.
}

  \label{fig:excess}
\end{figure}

Fig.~\ref{fig:excess} shows the $\gamma$-ray excess counts in the region around the SNR \g.
The image has been smoothed with a Gaussian kernel with standard deviation $0.06^{\circ}$.
The significance of the source detection amounts to 7.0 $\sigma$ using an integration radius of $0.1^{\circ}$.
The source centroid and extension were determined by fitting a 2D Gaussian, 
convolved with the H.E.S.S. PSF ($0.076^{\circ}$ for this analysis), to the unsmoothed $\gamma$-ray excess map.
The best-fit centroid is at $\alpha_{\mathrm{J}2000} = 18^{\mathrm{h}}18^{\mathrm{m}}3.4^{\mathrm{s}} \pm 4.6^{\mathrm{s}}_{\mathrm{stat}}$ and
$\delta_{\mathrm{J2000}} = -15^{\circ}27^{\prime}54^{\prime\prime} \pm 68^{\prime\prime}_{\mathrm{stat}}$ 
$(\mathrm{l} \sim 15.41^{\circ},\mathrm{b} \sim 0.17^{\circ})$, 
and the source is thus given the identifier \hessj. In addition to the statistics uncertainty in this fit, there is also a
systematic uncertainty due to the pointing precision of the telescope array (20$^{\prime\prime}$, \cite{gillessen}).
Both the best-fit position and the best-fit extension ($\sigma = 0.071^{\circ} \pm 0.015^{\circ}$, 68\% containment radius),
are indicated in Fig.~\ref{fig:excess}.

The results of this $\gamma$-ray morphological analysis are summarized in Table~\ref{tab:morphology}, 
together with updated values for the centroid position and average extension of the SNR \g\ in the radio. 
The latter have been estimated geometrically 
and the errors quoted represent the difference between the results for the 20\,cm image and the 90~cm image.
A comparison of these results shows that the centroids of \g\ and \hessj\ are compatible within 1-$\sigma$ errors. 
Furthermore, the extension of \hessj\ is significantly (at more than 2\,$\sigma$) smaller than the average extension of \g.

\begin{table}[h]
\centering
\begin{tabular}{ccc}
\hline
 & \hessj\ & \g\ \\
\hline
$\alpha_{\mathrm{J}2000}$ & $18^{\mathrm{h}}18^{\mathrm{m}}3.4^{\mathrm{s}} \pm 4.6^{\mathrm{s}}_{\mathrm{stat}}$ & $18^{\mathrm{h}}18^{\mathrm{m}}2^{\mathrm{s}} \pm 0.3^{\mathrm{s}}$ \\
$\delta_{\mathrm{J}2000}$ & $-15^{\circ}27^{\prime}54^{\prime\prime} \pm 68^{\prime\prime}_{\mathrm{stat}}$ & $-15^{\circ}28^{\prime}32^{\prime\prime} \pm 14^{\prime\prime}$ \\ 
Size & $8.5^{\prime} \pm 1.8^{\prime}_{\mathrm{stat}}$ & $(14.4^{\prime} \times 9.6^{\prime}) \pm 36^{\prime\prime}$ \\
\hline
\end{tabular}
\caption{Position and size of \hessj\ and SNR \g.}
\label{tab:morphology}
\end{table}

\subsection{Spectrum}
\begin{figure}[!t]
  \vspace{5mm}
  \centering
  \includegraphics[width=3.25in]{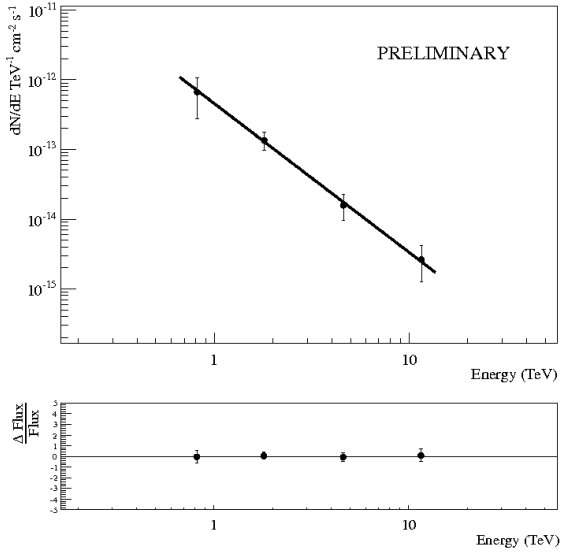}
  \caption{The top panel shows the differential energy spectrum of \hessj. The error bars represent 1-$\sigma$ statistical errors. The result of a power-law fit to the spectral data points is also shown. Events with energies between 0.6\,TeV and 13.5\,TeV were used in the fit, and the bottom panel shows the residuals.}
  \label{fig:spectrum}
\end{figure}

Spectral analysis was performed in a circular region centered on the best-fit position of \hessj\ and with a  
radius of 0.2$^{\circ}$, enclosing $\sim$90\% of the $\gamma$-ray emission. In this region, 116 excess $\gamma$-ray counts were found.
The differential spectrum, shown in Fig.~\ref{fig:spectrum}, is well fitted by a power law
$\mathrm{dN} / \mathrm{dE} = \phi_0 (\mathrm{E} / \mathrm{TeV})^{-1}$ with a photon index
$\Gamma = -2.1 \pm 0.3_{\mathrm{stat}} \pm 0.2_{\mathrm{syst}}$ and a flux normalisation at 1\,TeV of
$\phi_{\mathrm{0}} = (4.5 \pm 1.5_{\mathrm{stat}} \pm 0.9_{\mathrm{syst}}) \times 10^{-13}$\,cm$^{-2}$\,s$^{-1}$\,TeV$^{-1}$.
This corresponds to an integral flux $F(>1$\,TeV$)=(4.0 \pm 1.6_{\mathrm{stat}} \pm 0.8_{\mathrm{syst}}) \times 10^{-13}$\,cm$^{-2}$\,s$^{-1}$,
equivalent to $\sim$1.8\% of the flux of the Crab Nebula in the same energy range.

\section{\g: A new composite SNR?}
\begin{figure*}[!t]
  \centerline{\includegraphics[width=3.25in]{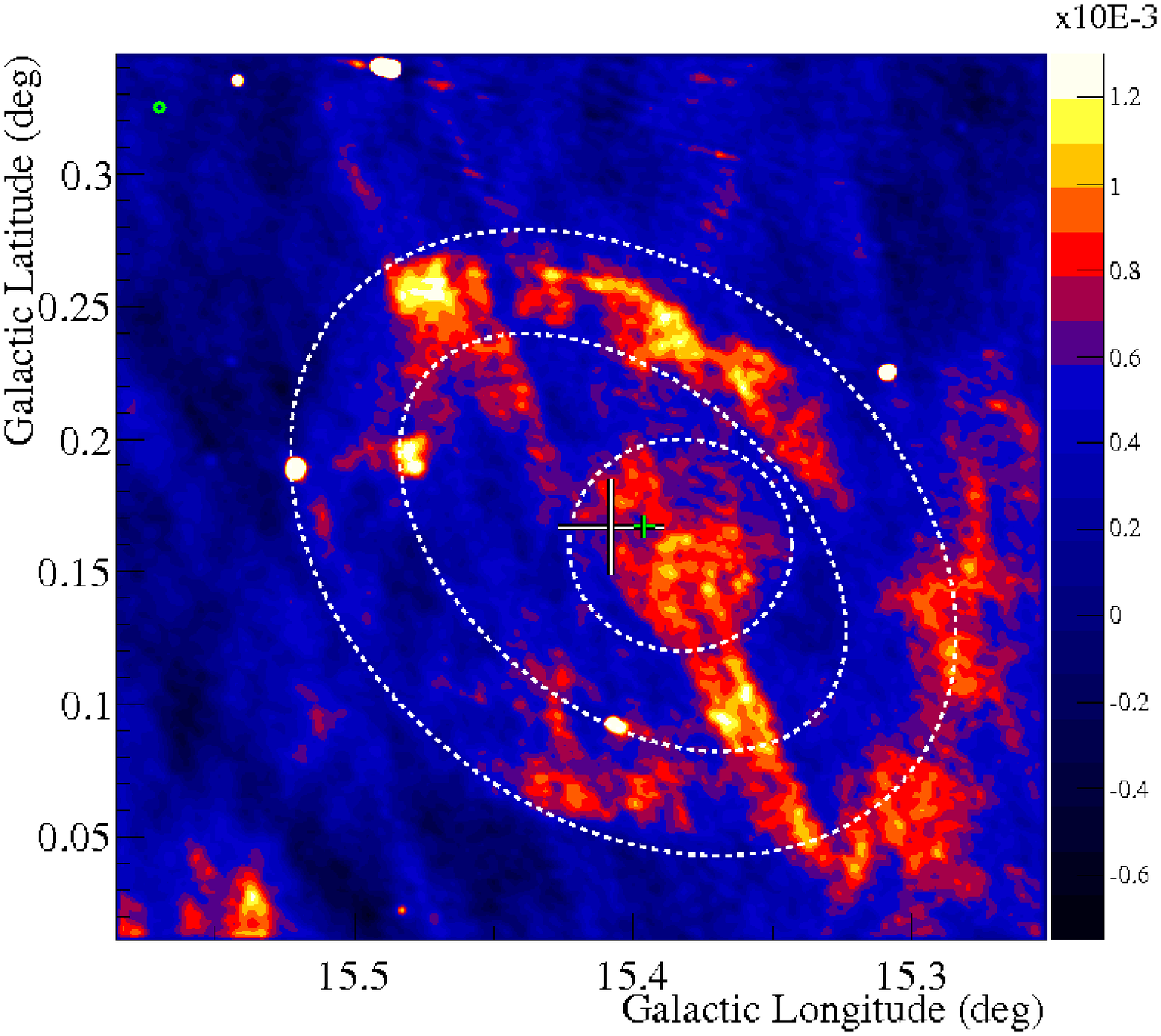}
    \hfil
    \includegraphics[width=3.25in]{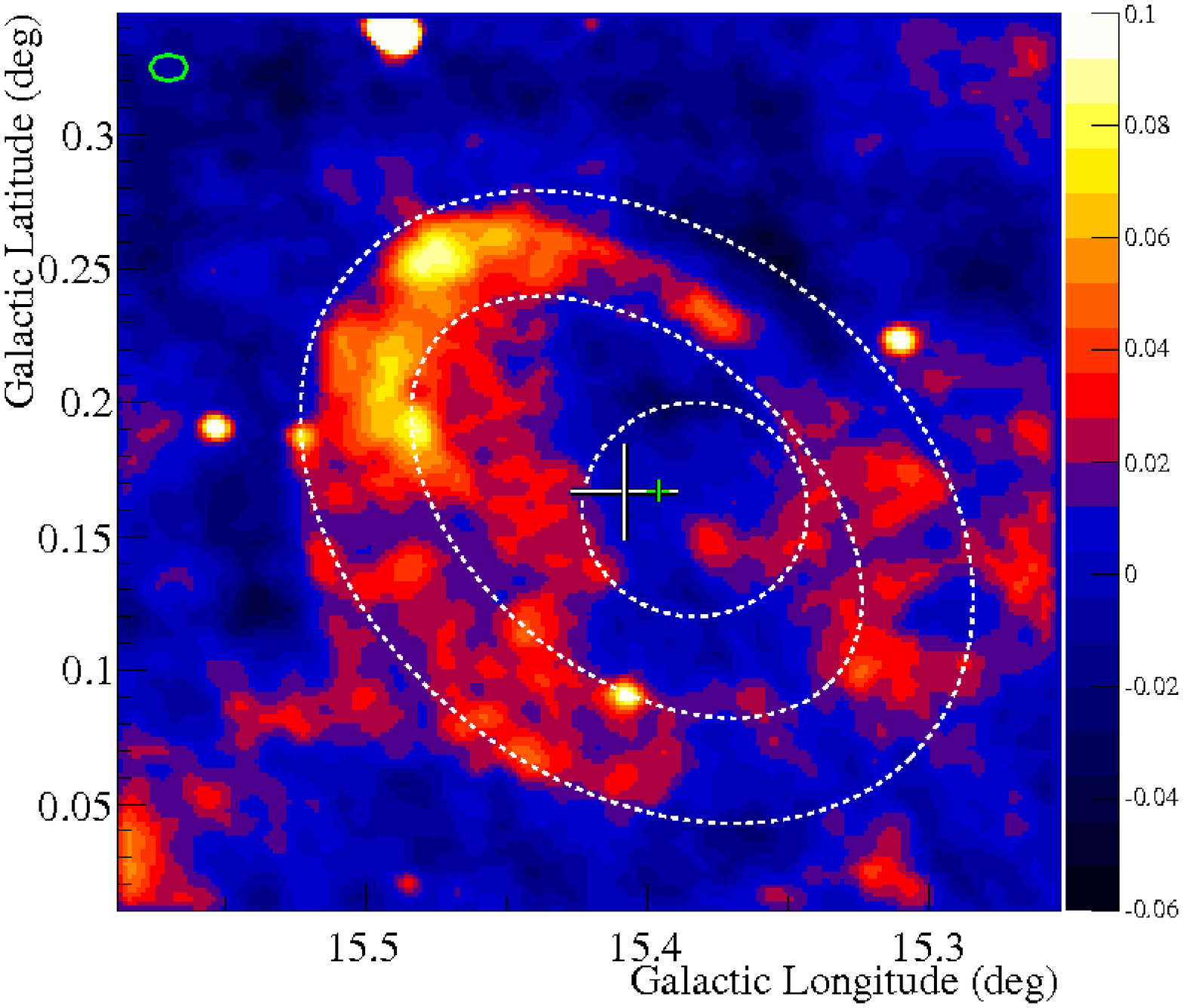}}
  \caption{
Images of VLA 20-cm (left) and 90-cm (right) radio observations of \g\ in units of Jy\,beam$^{-1}$. 
The integration regions are indicated by white ellipses and a circle, respectively. 
The centroids of \hessj\ and \g\ are indicated by a white and green cross. 
The extension of the cross corresponds to the difference in the estimated centroid positions for the 20-cm and 90-cm data.
The images are shown in their original beam sizes, 
the size of which is indicated by the small green ellipses in the upper left corners.
}
  \label{fig:radio}
\end{figure*}

Since the discovered $\gamma$-ray emission is found to be concentrated in the centre of the SNR \g\ 
it is natural to consider \hessj\ as a candidate TeV PWN associated with the SNR.
In this scenario, the PWN may also be visible in the radio. 
Archival radio data from MAGPIS \cite{magpis} was used to investigate this hypothesis. 
Observations of the region around \g\ are available at both 20\,cm and 90\,cm wavelengths, 
where the PSF (here, half-power beamwidth) of the observations are $6.2^{\prime\prime}\times5.4^{\prime\prime}$ and 
$24^{\prime\prime}\times18^{\prime\prime}$, respectively. 
The radio intensities observed in the field-of-view of \g\ at these two wavelengths 
are shown in Fig.~\ref{fig:radio}.
The SNR has a clear shell structure at both wavelengths, 
but notable differences in morphology, in particular in the central region.
While the SNR centre appears largely void of radio emission at 90~cm, hints of a nebula with a diameter of
$\sim$$5^{\prime}$ are visible at 20~cm. This suggests there could be a varying spectral index across the SNR and putative PWN.
A closer inspection of the 20-cm image reveals a strong source in the far north of the observation
that may have produced spurious emission and image artefacts.
The radio analyses in this proceeding should therefore be considered preliminary.


With these caveats noted,
the integrated flux densities for both the central and shell components of SNR \g\ were determined. 
The images were first convolved to a common beam size of $25^{\prime\prime} \times 25^{\prime\prime}$ and
point sources\footnote{From SIMBAD, \url{http://simbad.u-strasbg.fr/simbad}},
were removed by masking areas 1.5 times the beam size. 
The RMS noise was then derived from each (convolved) image and were found to be $\sigma_{90} = 20$\,mJy\,beam$^{-1}$ and
$\sigma_{20} = 4$\,mJy\,beam$^{-1}$, respectively. 
Integrating within the shell and central regions, denoted by the (white) ellipses and circle in Fig.~\ref{fig:radio},
flux densities were measured and are given in Table~\ref{tab:radioflux}. 
Uncertainties in the flux density include contributions from i) image noise, ii) calibration, and iii) wrong zero levels, 
as suggested in \cite{radiofluxerrors}. From the flux densities derived at 20- and 90-cm wavelengths, spectral indicies could be derived
for the two regions; they are given in Table~\ref{tab:radioflux}. 
The shell exhibits a steep spectrum ($\alpha \sim -0.7$), in agreement with \cite{brogan2006} and indicative 
non-thermal synchrotron emission. 
However, the central region is found to exhibit a nearly flat spectrum ($\alpha \sim 0.3$), supporting the central PWN
hypothesis \cite{pwnevolution}, and further adding to the case that \g\ may actually be a composite SNR. 

\begin{table}[h]
\centering
\begin{tabular}{cccc}
\hline
Region & Flux (90\,cm) & Flux (20\,cm) & Index \\
\hline
Centre & ($0.8 \pm 0.2$)\,Jy & ($1.2 \pm 0.2$)\,Jy & $0.3 \pm 0.2$ \\
Shell & ($14.0 \pm 1.0$)\,Jy & ($4.6 \pm 0.5$)\,Jy & $-0.7 \pm 0.1$ \\
\hline
\end{tabular}
\caption{Derived flux densities and spectral indicies ($S \propto \nu^{\alpha}$) for the central and shell region of \g.}
\label{tab:radioflux}
\end{table}

\section{Discussion and conclusions}
Since a chance positional coincidence of the centroid of the VHE $\gamma$-ray emission
with the center of the radio shell of \g\ seems highly unlikely, 
an association between \hessj\ and the SNR is assumed here.  
The question is then whether the VHE emission originates from the shell of the SNR 
or from a putative PWN embedded inside the SNR. 
However, since the extension of the H.E.S.S. source is significantly smaller than the extension of the SNR as seen in the archival
VLA data, it is clear that the bulk of the VHE emission originates 
from a yet unknown source located within the shell. 
The discovery of a radio or X-ray PWN counterpart to \hessj\ would confirm this scenario.
High-quality X-ray observations of the region are unfortunately lacking, 
but preliminary analyses of 20-cm and 90-cm radio observations 
suggest that the SNR is composed of two distinct regions with different spectra. 
The central region is found to be characterised by a relatively flat radio spectrum, 
characteristic of the central PWN of a composite SNR \cite{pwnevolution}, while the shell is characterised
by a steeper spectrum, indicative of the non-thermal synchrotron emission expected from an SNR shell.
\g\ thus appears to be an example of a composite SNR, similar to G\,0.9$+$0.1, the first composite SNR to be detected at 
TeV energies. If the nature of \g\ can be confirmed, \hessj\ would be the first case in which a SNR is identified as a
composite SNR on the basis of VHE $\gamma$-ray observations.

Additional supporting evidence for a PWN origin of the TeV emission would be
the detection of a pulsar within the central regions of the SNR. 
Unfortunately, no pulsar is currently known, in radio or X-rays, inside the shell of \g, nor are there any known hard X-ray 
point sources. However, a typical young radio pulsar located at $\sim$9\,kpc with an assumed luminosity
L$_{21\mathrm{\,cm}} \sim 56$\,mJy\,kpc$^2$
(i.e. similar to the median of high-luminosity, young, rotation-powered pulsars estimated by \cite{camilio})
would have a flux density of S$_{21\mathrm{\,cm}} \sim 0.08$\,mJy, 
considerably lower than the sensitivity of the current VLA data. 
Beaming effects could also make the pulsar virtually undetectable in radio. 
If future observations of \g\ are able to detect the putative central pulsar associated with \hessj,
it is expected to have a very high spin-down luminosity ($\dot{\mathrm{E}} \sim 4 \times 10^{37}$\,erg\,s,
as seen in other TeV $\gamma$-ray emitting PWNe \cite{radiog0.9}.

Little is known about the distance and age of \g. Using the uncertain $\Sigma-D$ relationship \cite{sigmaD}, 
a distance of $10 \pm 3$\,kpc is derived.
Adopting 10\,kpc as the nominal distance of the SNR would imply a Sedov age of $\sim$9\,kyr,
assuming an ambient ISM density of 1\,cm$^{-3}$.
However, varying the density by $\pm$50\% and taking into account uncertainties in the $\Sigma-D$ relationship, 
an age as low as 4\,kyr can be derived. These age estimates would make SNR \g\ older than similar systems previously
detected in VHE $\gamma$-rays.

\section{Acknowledgments}
The support of the Namibian authorities and of the University of Namibia in facilitating the construction and operation of H.E.S.S. is gratefully acknowledged, as is the support by the German Ministry for Education and Research (BMBF), the Max Planck Society, the French Ministry for Research, the CNRS-IN2P3 and the Astroparticle Interdisciplinary Programme of the CNRS, the U.K. Science and Technology Facilities Council (STFC), the IPNP of the Charles University, the Polish Ministry of Science and Higher Education, the South African Department of Science and Technology and National Research Foundation, and by the University of Namibia. We appreciate the excellent work of the technical support staff in Berlin, Durham, Hamburg, Heidelberg, Palaiseau, Paris, Saclay, and in Namibia in the construction and operation of the equipment.



\clearpage

\end{document}